\documentclass{vgtc}                          

\usepackage{mathptmx}
\usepackage{graphicx}
\usepackage{times}
 \usepackage{amsmath}

\vgtccategory{Research}

\vgtcinsertpkg

\title{Crowdsourcing Autonomous Traffic Simulation}
\author{Hua Wang \quad Wenshan Zhao \quad Zhigang Deng \quad Mingliang Xu}

\abstract{
We present an innovative framework, Crowdsourcing Autonomous Traffic Simulation (CATS) framework, to safely implement and realize orderly traffic flows. We first provide a semantic description of the CATS framework using theories of economics to construct coupling constraints among drivers, in which drivers monitor each other by making use of transportation resources and driving credit. We then introduce an emotion-based traffic simulation, which utilizes the Weber-Fechner law to integrate economic factors into drivers' behaviors. Simulation results show that the CATS framework can significantly reduce traffic accidents and improve urban traffic conditions.

}


\begin{document}
\maketitle
\numberwithin{equation}{section}
\section{INTRODUCTION}
Rising traffic problems, such as congestion, accidents, and violations amongst others, are inevitable in large and growing metropolitan areas across the world. A significant portion of the population is plagued by traffic related problems every day. The issue of effectively managing traffic has become a mounting challenge for numerous cities.

Numerous methods have been used to address urgent traffic problems, such as number plate based restrictions, advocating the use of public transit, and traffic camera monitoring systems amongst others\cite{zhu2013rationing,shi2014optimization}. In most of these mentioned methods, traffic participants (drivers for example) are subject to law enforcement measures. Most of them are generally egocentric and prefer to take chances by committing traffic violations in the absence of supervision and enforcement, which is a kind of non-cooperative game that occurs in the real world. As a result, there have been no significant improvements of urban traffic conditions in large cities using the above methods\cite{Laurance,Perz}.

 This paper proposes a new traffic management system called Crowdsourcing Autonomous Traffic Simulation (CATS) framework to tackle the above-mentioned problems. In this system, we present the following views based on theories of economics and psychology:
 (1) A traffic camera is present in each vehicle to achieve the aim of mutual supervision. Each driver possesses dual identities as a ``law enforcer'' and ``object of law-enforcement'' at the same time.
 (2) Drivers would be punished after violating traffic laws. In addition, drivers would be able to obtain ``rewards'' by reporting violations of other drivers (rewards are to be covered by fines collected from perpetrators of traffic violations in the form of transportation resource value).
 (3) In addition to receiving economic punishments, the perpetrator will also be deducted a certain amount of driving credit. Credit can be obtained on a regular basis when the system distributes equal amounts to all drivers, and it has a validity period. The credit value of each driver is limited within a certain time period. Drivers are forbidden to drive after credit has been deducted completely. This driving credit regime serves as a deterrence for traffic misbehaviors and non-compliance.

\begin{figure*}[t]
\centerline{\includegraphics[width=6.0 in]{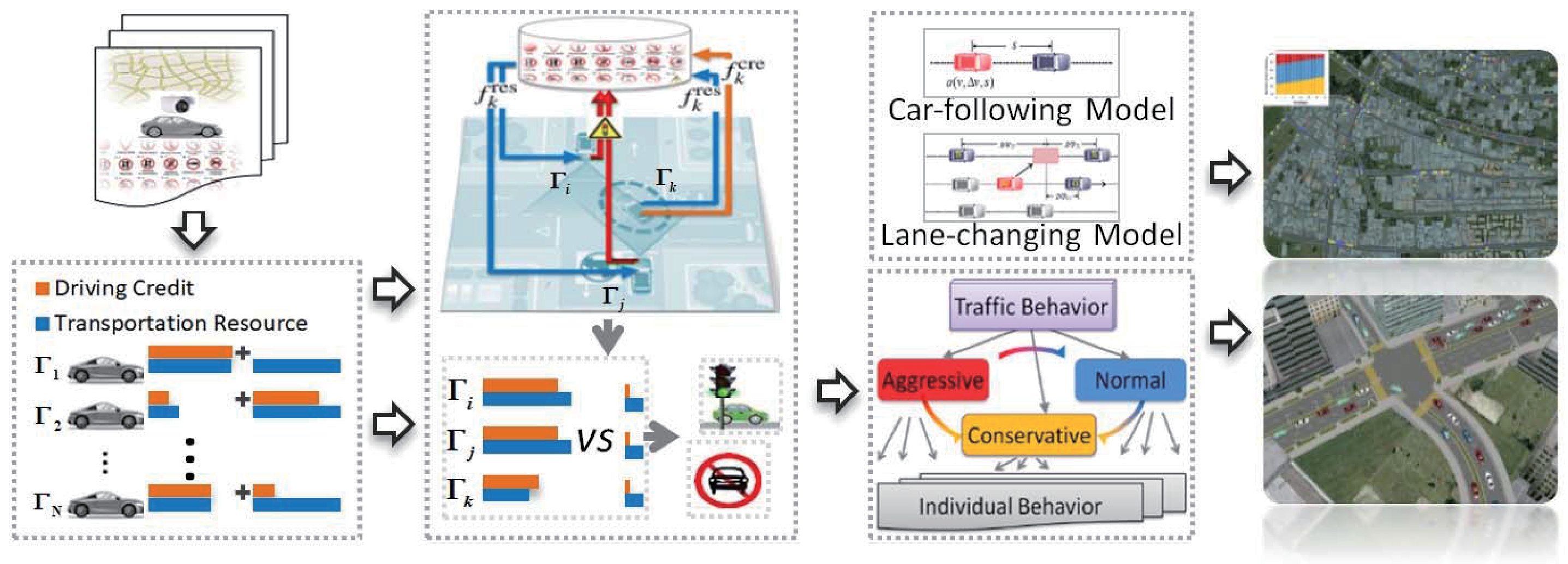}}
\caption{\label{fig-framework} The framework of our CATS.}
\end{figure*}

In order to develop the above-mentioned CATS framework, we use a multi-agent method to simulate the evolution of vehicle flow dynamics within the system. Specifically, we firstly use a car-following model and a lane-changing model to model vehicles' movements. We then model psychological dynamic of drivers in the CATS framework by the Weber-Fechner law to impact vehicles' movements, thus driving traffic simulations in the CATS framework (Figure \ref{fig-framework}).

Our contributions are as follows:
\begin{itemize}
  \item Constructing coupling constraints among drivers: We consider drivers as law enforcers and objects of law-enforcement at the same time and enforce reward and punishment mechanisms to monitor each other.
  \item	Modeling the relationship between the above coupling constraints and drivers' behaviors: We use the Weber-Fechner law to measure drivers' sensations in relation to the physical magnitude of reward and punishment stimuli.
  \item	Introducing a multi-agent traffic simulation method to develop the CATS framework: We introduce a multi-agent traffic simulation method to model vehicles' movements by combining traffic economics(reward and punishment mechanisms), psychology (driving behaviors) and dynamics.
\end{itemize}

\section{RELATED WORK}

In this section, we firstly give some descriptions about existing methods for traffic management. We then describe the latest studies about traffic simulations as we drive traffic simulations in the CATS framework.
\subsection{Traffic Management}

There are about two classes of traffic management methods: road space rationing and economic punishment.

Road space rationing is aimed to reduce the negative externalities generated by urban travel demand in excess of available supply or road capacity, through artificially restricting demand by rationing the scarce common good road capacity. A number of countries have implemented road space rationing such as allowing travel only on alternate-days, introducing driving restrictions\cite{han2010efficiency}, no-drive days and advocating the use of public transit over many years.
However, there have been no significant improvements in urban traffic conditions in large cities. The limited transportation resources of cities are unable to meet the continuously increasing traffic demands during the process of urbanization. The experience of developed countries has proved that it is difficult to fundamentally solve these increasingly serious traffic problems by simply increasing the supply of transport resources. It is easy to fall into the following vicious cycle: congestion $\rightarrow$ constructing new roads $\rightarrow$ relief $\rightarrow$ inducing new demand $\rightarrow$ congestion again\cite{Laurance,Perz}.

From a behavioral science perspective, economic punishment is one of the most direct and effective approaches to ease urban traffic congestion. Adam et al. proposed a solution to guide and regulate urban traffic by levying road congestion fees, and thus being able to alleviate traffic congestion\cite{Adam,Fosgerau}. However, such solutions not only increase the cost of driving, but also give rise to public concerns over the  redistribution of the collected fees \cite{Small,Adam}. To offset the negative effects of congestion fees and ensure that people are not discouraged to move around freely, Kockelman et al. presented a credit-based congestion pricing strategy. The strategy associates with driving under congested conditions in which road tolls are based on the negative externalities. Generated tolls are returned to all licensed drivers in a uniform fashion \cite{Kockelman}. Verhoef et al. explored the possibilities of using tradeable permits in the regulation of road transport externalities  \cite{Verhoef}. Akamatsu presented a transportation demand management scheme called ``tradable bottleneck permits'' \cite{Akamatsu}. Yang et al. examined an alternative simple credit distribution and charging scheme. In this scheme, governments initially issue credit to all eligible travelers. The credit can then be traded freely in a competitive market without government intervention \cite{Yang2011}. Charges regulate traffic flow as pricing does in both static and dynamic control settings by allowing for unrestricted trading of credit. The CATS framework is inspired by this scheme. In the above economic punishment measures, traffic cameras play a vital role in the enforcement of traffic laws. However,  a number of security risks are associated with this \cite{Langland-Orban}. Research shows that traffic camera systems could increase the number of accidents, such as rear-ending and hurting pedestrians at certain locations. The risk of traffic accidents exists during the final seconds of a green light traffic signal. Moreover, when drivers are aware of the locations of traffic cameras, the effectiveness of traffic camera systems is limited \cite{Hu,McCartt,Pulugurtha}.

Traffic problems appear to be caused by a spatial and temporal mismatch of the supply and demand of transportation resources. When considering the issue from the perspectives of economics and psychology, traffic problems are primary as a result of conflicts caused by traffic participants, who are generally egocentric and prefer to take chances by committing traffic violations in the absence of supervision and traffic law enforcement. This could be referred to as ``non-cooperative game'' that occurs in the real world.

\subsection{Traffic Simulation}
Traffic simulation models could be mainly classified into two broad categories based on the level of simulation details \cite{chao2019a,xu2014crowd}:  macroscopic and microscopic models. Macroscopic models formulate the relationships among traffic flow characteristics such as density, flow, and mean speed of a traffic stream. These include anisotropic models\cite{Aw,Jiang,Zhang2002} and lattice hydrodynamic models \cite{Ge,Peng}. Microscopic models describe traffic at a high level of detail (agent level) and attempt to model the actions and reactions of vehicles as accurately as possible. These include optimal velocity models \cite{Helbing,Nagatani}, intelligent driver models \cite{Treiber}, and the cellular automaton models \cite{Belitsky}.

In order to meet actual demands in computer graphics, Sewall et al. first introduced the concept of ``virtual traffic'' and used the A* algorithm to achieve virtual traffic reconstruction. Every vehicle in the model should consider all vehicles in the scenario\cite{Sewall2011a}. Sewall et al. also proposed a hybrid technique by coupling continuum and agent-based traffic models\cite{Sewall2011b}. Wilkie et al. introduced a fast technique to reconstruct traffic flows from in-road sensor measurements or procedurally generated data for interactive 3D graphics applications\cite{Wilkie}. Wang et al. developed the traditional lattice hydrodynamic model by an interaction term to model traffic interactions between lanes\cite{Wang2014b}. Shen et al. presented a new agent-based system for detailed traffic animation on urban arterial networks with diverse junctions such as signalized crossing, merging and weaving areas\cite{Shen}. Yang et al. presented a novel traffic animation method to promote the sense of immersion of the virtual traffic flow by inserting virtual vehicles into the real trajectory data based on the intelligent driver model\cite{Yang2015}. Ignacio provided an interactive approach which enables a designer to specify a desired vehicular traffic behavior\cite{GarciaDorado}. Wang et al. used shadow traffic to model
abnormal traffic behavior\cite{wang2018shadow}. Li et al. presented a city-scale traffic animation using statistical learning and metamodel-based optimization\cite{li2017city-scale}. Chao et al.introduced a novel data-driven scheme to generate a large set of new vehicle trajectories through the non-trivial fusion of texture synthesis and trafﬁc behavior rules\cite{chao2018realistic}.

The significant majority of existing traffic simulation models rarely gives any consideration to economic factors. Parameters in these models are determined by experience and some observations. Therefore, these models are unable to provide accurate descriptions of enormous and complex transportation systems.

\section{MODELING OF THE CATS FRAMEWORK}

A multi-agent method is used to abstract the CATS framework as a generic 5-element tuple $\Theta = \left( {\Gamma ,\Omega ,\Phi ,\Lambda ,\bar \Lambda } \right)$, where $\Theta$ represents the ``driver-vehicle-road'' traffic system; $\Gamma$ represents the ``driver-vehicle'' unit set; $\Omega$  represents the set of rules based on the CATS framework; $\Phi$ represents the traffic resources provided by the system i.e. road space available for driving; $\Lambda$ represents the types of traffic violations identified by the system and the corresponding penalties; and $\bar \Lambda$  represents the type of spontaneous comity behaviors identified by the system and the corresponding rewards.

{\it Definition 1} : 
${\Gamma _i} \in \Gamma $ is the ``driver-vehicle'' integrated agent. Here, it is referred to as dri-vehicle without discrimination. In addition, $N = \left| \Gamma  \right|$ represents number of dri-vehicles moving on roads.

{\it Definition 2} : 
 The transportation resource value  ${\wp _i}\left( t \right)$ represents the amount of transportation resource that $\Gamma _i $ owns at the time of $t$.

{\it Definition 3} : 
The driving credit value ${\ell _i}\left( t \right)$ represents the amount of credit  that $\Gamma _i $ owns at the time of $t$.

{\it Definition 4} : 
Denote the set of dri-vehicles with violations that are being detected by the camera on ${\Gamma _i}$ at the time $t$ as ${A_i}\left( t \right)$. For ${\Gamma _j} \in {A_i}\left( t \right)$, denote the set of dri-vehicles that are detecting the violation of $\Gamma _j$ at the time of $t$ as ${\Psi_j}\left( t \right)$.

Based on the above definitions, the following sets of assumptions are detailed.

{\it Hypothesis 1} : 
During system initialization, the system allocates a reasonable amount of transportation resources ${\wp _0}$ equally to every dri-vehicle free of charge. Then, for each ${\Gamma _i}$, ${\wp _i}(0) = {\wp _0}$, $\wp _0$ represents the initial purchase power of ${\Gamma _i}$. After a period of $\Delta T_1$, every dri-vehicle is given an equal amount of transportation resource $\wp _0$, which has no time validity limit and can therefore be accumulated. That is, when $t = n\Delta T_1$ $\left( {n \in N^+ } \right)$, then ${\wp _i}\left( t \right) = {\wp _i}\left( t \right) + {\wp _0}$. It is agreed that during a period of $\left[ {n\Delta T_1,\left( {n + 1} \right)\Delta T_1} \right]$, $n \in N$, if the transportation resource value of ${\Gamma _i}$, ${\wp _i}\left( t \right) \le {\wp ^{\min }}$, ${\wp ^{\min }} \in \left( {0,{\wp _0}} \right)$, then the dri-vehicle ${\Gamma _i}$ is prevented from making use of any road resources before transportation resources have been received again.

{\it Hypothesis 2} : 
 During system initialization, the system allocates an equal amount of driving credit ${\ell _0}$ free of charge. Then, for ${\Gamma _i}$, ${\ell _i}(0) \equiv {\ell _0}$. Every time following a period of $\Delta T_2$, each dri-vehicle will be given an equal amount of driving credit ${\ell _0}$ again, which hold a time validity limit and cannot be traded freely. When ${t_0} = n \Delta T_2, n \in N$ , the driving credit value of  ${\Gamma _i}$ is ${\ell _i}({t_0}) \equiv {\ell _0}$ . It is agreed that during a period of $\left[ {n\Delta T_2,\left( {n + 1} \right)\Delta T_2} \right]$ , the upper limit of accumulated driving credit of ${\Gamma _i}$ is ${\ell _0}$ .When $t \in\left[ {n\Delta T_2,\left( {n + 1} \right)\Delta T_2} \right]$ and ${\ell _i}\left( t \right) \le 0$ , ${\Gamma _i}$ will not be able to make use of any road resources during the subsequent period of $ \left[ {t,\left( {n + 1} \right)\Delta T_2} \right]$.

\begin{figure}[h]
\centerline{\includegraphics[width=1.5in]{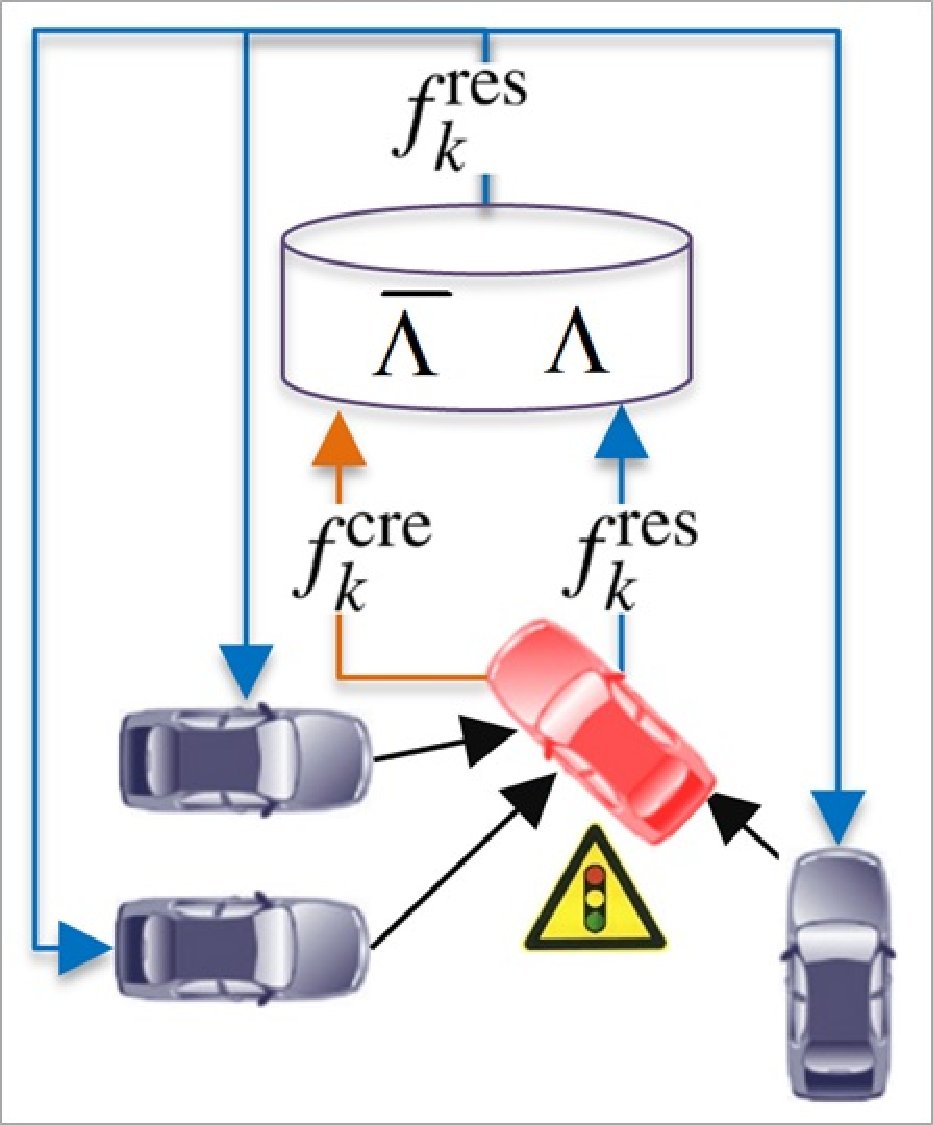}}
\caption{\label{fig-hoat} The reward and punishment system among dri-vehicles for a traffic violation. }
\end{figure}

{\it Hypothesis 3} : 
$\forall {\Gamma _i} \in \Gamma $ is able to detect all misbehaving dri-vehicles within the detection scope. Meanwhile, the misbehavior of ${\Gamma _i}$ can also be detected by surrounding dri-vehicles. The set of misbehaving dri-vehicles detected is denoted by ${\Gamma _i}$ at the time of $t$ as ${A_i}(t)$, then for $\forall {\Gamma _j} \in {A_i}(t)$, it will not only be deducted a certain amount of driving credit, but will also be required to pay a fine immediately to  ${\Gamma _i}$ in the form of transportation resource value. The actual amount of fine is jointly determined by ${\Gamma _j}$'s violation type $k \in \Lambda$, and $\left| {{\Psi _j}\left( t \right)} \right|$, which is the number of vehicles that detected the violation of ${\Gamma _i}$ at the time of $t$. The amount of driving credit to be deducted as a result of violation of type $k$ is denoted as $f_k^{\mathrm{cre}}$, and the amount of transportation resource value to be deducted are denoted as $f_k^{\mathrm{res}}$. Then, the amount of the fine ${\Gamma _j}$   needs to pay ${\Gamma _i}$   at the time of $t$, can be defined as ${f_{j \to i}}\left( t \right) = {{f_k^{\mathrm{res}}} \mathord{\left/
 {\vphantom {{f_k^{\mathrm{res}}} {\left| {{\Psi _j}(t)} \right|}}} \right.
 \kern-\nulldelimiterspace} {\left| {{\Psi _j}(t)} \right|}}$ . The dri-vehicle which has violated traffic laws will be notified of the penalty immediately (Figure \ref{fig-hoat}).

{\it Hypothesis 4} : 
$\forall {\Gamma _i} \in \Gamma $  can obtain information in relation to currently congested areas $J = \{ are{a^{\mathrm{jam}}}\left( t \right)\} $.
 When  ${\Gamma _i}$ chooses to travel to a location within a congested area $are{a^{\mathrm{jam}}}\left( t \right) \in J$ , it will be charged a congestion fee in the form of traffic resource value ${\wp _i}\left( t \right)$.

By taking into account the above mentioned assumptions and traffic psychology theories, which describe that drivers' behaviors are affected by factors such as emotions, next we will show how vehicles move within the CATS framework.

\section{CATS SIMULATION}

\subsection{Preliminaries}

\subsubsection{Car-following Model}
In the CATS framework, drivers adjust their driving behaviors according to their driving credit values. However, they also comply with car-following rules. Here we adopt the intelligent driving model(IDM)\cite{Treiber}.

\[a\left( {v,\Delta v,s} \right) = {a_{\max }}\left[ {1 - {{\left( {\frac{v}{{{v_0}}}} \right)}^\delta } - {{\left( {\frac{{{s^*}\left( {v,\Delta v} \right)}}{s}} \right)}^2}} \right]\]

Here, $a$ denotes the accelerated velocity, $v$ is the velocity, $s$ is the distance between two vehicles, $\Delta v$ is the velocity difference, ${s^*}\left( {v,\Delta v} \right) = {s_0} + Tv + \frac{{v\Delta v}}{2\sqrt {a_{\max }{b_{\mathrm{com}}}}}$ is the desired minimum gap. $T$ represents the safe time headway. ${a_{\max }}$ is the maximum acceleration. ${b_{\mathrm{com}}}$ is the desired deceleration. $\delta$ is the acceleration exponent and $s_0$ is the jam distance.
\subsubsection{Lane-changing Model}

Lane changes are common behaviors which can be observed in various traffic scenarios such as when vehicles travel on on/off-ramps, during lane closings, interactions, at interchanges, and within traffic accident segments. Modeling lane changes naturally play an important role in traffic simulations and animations to accurately represent real-world situation and dynamic realism \cite{Wang2014a}.

There are two steps to determine a lane-changing process: lane-changing necessities and lane-changing feasibilities.
First, a vehicle possesses a lane-changing necessity if one of the following conditions is applicable:
\begin{itemize}
  \item The speed of its leader vehicle in the current lane is slower than that of its leader vehicles in neighboring lanes (the ratio of which is $\eta$ and $\eta<1$ ).
  \item	The lane is being closed as a result of lane merging and traffic accidents.
  \item	Ramp terminal scenarios.
\end{itemize}

Drivers judge feasibilities of lane-changing maneuvers according to surrounding conditions. The surrounding conditions describe that there should be sufficient gaps between the subject vehicle (S) and surrounding vehicles to avoid collisions during lane changes. S, which is following the leader vehicle (CL) in the current lane, wishes to merge into its target lane, in between the leader vehicle (TL) and the follower vehicle (TF). The gaps between s and CL (TL / TF) should be larger than some given smallest acceptance gaps during the lane-changing process to ensure collisions are avoided.

Based on some existing lane-changing models\cite{Hidas, Wang2014a}, three gaps are considered here. Let the gap between $\mathrm{S}$ and $\mathrm{CL}$ at the end of the lane-changing process be $gap_{\mathrm{CL}}$, the gap between $\mathrm{S}$ and $\mathrm{TL}$ at the end of the process be $gap_{\mathrm{TL}}$, the gap between $\mathrm{TF}$ and $\mathrm{V}$ at the end of the process be $gap_{\mathrm{TF}}$ (Figure \ref{fig-lane-changing}). In order to avoid lane-changing conflicts, $gap_{\mathrm{CL}}$, $gap_{\mathrm{TL}}$ and $gap_{\mathrm{TF}}$ should satisfy the following formulas:
\begin{figure}[h]
\centerline{\includegraphics[width=3.0in]{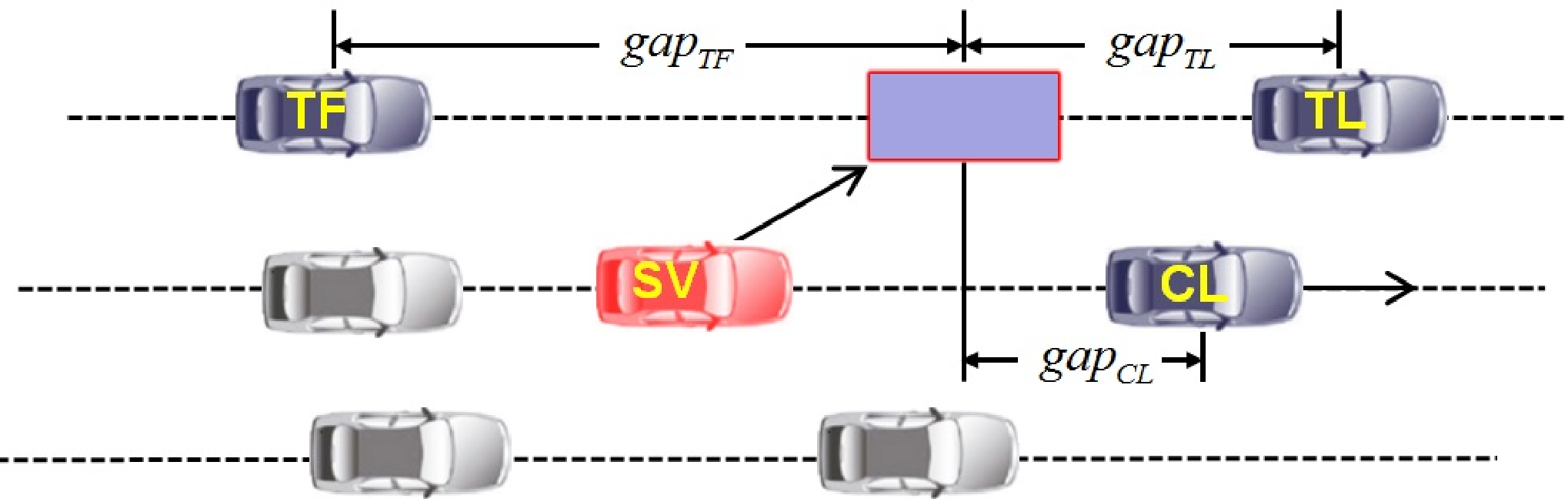}}
\caption{\label{fig-lane-changing} Vehicles that produce effects on a lane-changing process for $\mathrm{SV}$. The red vehicle is $\mathrm{SV}$. It decides to change its lane to the left. The red box is a clone of $\mathrm{SV}$ to describe a suppositional position of $\mathrm{SV}$ after it finishes the desirable lane-changing process.}
\end{figure}

\[gap_{\mathrm{CL}}\geq gap_{\mathrm{CL,min}}\]
\[gap_{\mathrm{TL}}\geq gap_{\mathrm{TL,min}}\]
\[gap_{\mathrm{TF}}\geq gap_{\mathrm{TF,min}}\]

$gap_{\mathrm{TL,min}}$ and $gap_{\mathrm{TF,min}}$ are the same as Hidas' model \cite{Hidas}, which are functions of velocities. We let the expression of $gap_{\mathrm{CL,min}}$, which is a function of  the velocity of  $\mathrm{CL}$, be $a*gap_{\mathrm{TL,min}}$. $a$ is constant and less than 1. $gap_{\mathrm{TL,min}}$, $gap_{\mathrm{TF,min}}$ and $a$ are restricted by a number of factors, such as weathers, driving behaviors and pavement conditions etc.

Let S, CL, TL, and TF travel at even speeds and $gap_{\mathrm{TL,min}}$, $gap_{\mathrm{TF,min}}$, $gap_{\mathrm{CL,min}}$  be a linear relation to velocities. Let $gap_{\mathrm{TL,min}} = {\mu _{1}}{v_{\mathrm{S}}}$,
$gap_{\mathrm{TF,min}} = {\mu _{2}}{v_{\mathrm{S}}}$,
$gap_{\mathrm{CL,min}} = {\mu _{3}}{v_{\mathrm{S}}}$, in which $\mu _{_1}$ , $\mu _{_2}$ , $\mu _{_3}$   are constants, then

\[{v_{\mathrm{S}}}\Delta \tau  \le {v_{\mathrm{CL}}}\Delta \tau  + {gap_{\mathrm{CL,0}}} - {\mu _{1}}{v_{\mathrm{S}}}\]
\[{v_{\mathrm{S}}}\Delta \tau  \le {v_{\mathrm{TL}}}\Delta \tau  + {gap_{\mathrm{TL,0}}} - {\mu _{2}}{v_{\mathrm{S}}}\]
\[{v_{\mathrm{S}}}\Delta \tau  \le {v_{\mathrm{TF}}}\Delta \tau  + {gap_{\mathrm{TF,0}}} - {\mu _{3}}{v_{\mathrm{S}}}\]

 Here $\Delta\tau $ specifies the lane-changing duration. $v_{\mathrm{SV}}$, $v_{\mathrm{CL}}$,  $v_{\mathrm{TL}}$, $v_{\mathrm{TF}}$ are the velocities of S, CL, TL, TF, respectively, before they change lanes.  $gap_{\mathrm{CL,0}}$, $gap_{\mathrm{TL,0}}$, $gap_{\mathrm{TF,0}}$ are gaps between S and CL, S and TL, S and TF, respectively.

\subsubsection{Weber-Fechner Law}

Weber-Fechner law is used to express the relationship between psychological and physical quantity. It can be expressed by the following formula:

\[{\kappa _1} = \varepsilon \ln {\kappa _2}\]

Where ${\kappa _1}$  represents sensory intensity, and ${\kappa _2}$  represents stimuli intensity.

The law states that psychological quantity is a logarithmic function of the stimuli. That is, sensory intensity increases in an arithmetic progression when stimuli intensity increases in a geometric progression. This law is widely used in psychology, acoustics, and marketing. Using marketing as an example, the Weber-Fechner law dictates that the buyer's perception of price is more dependent on the percentage of change than the absolute value of the change.

Drivers tend to adopt a conservative driving manner when traffic resources and driving credit are low. However, when traffic resources and driving credit are relatively abundant, the impact of one violation deduction unit is smaller than when the values are low. This shows that there is a Weber-Fechner law relationship between the driver's credit value and his driving behavior.

\subsection{CATS Simulation based on the Weber-Fechner Law}

The movement of a dri-vehicle in the CATS simulation is based on the car-following model and lane-changing model presented in Section 4.1. In the CATS simulation, the parameter vector $\left( {{v_0}, T, {a_{\max }}, {b_{com}}, \delta , {s_0}, \eta, {\mu _1}, {\mu _2}, {\mu _3}, \Delta \tau} \right)$ of the car-following model and lane-changing model is affected by the driving credit and road resources. By manipulating the values of car-following and lane-changing model parameters, the rules of the CATS simulation influence the movement of dri-vehicles.

\begin{figure}[h]
\centerline{\includegraphics[width=3.3in]{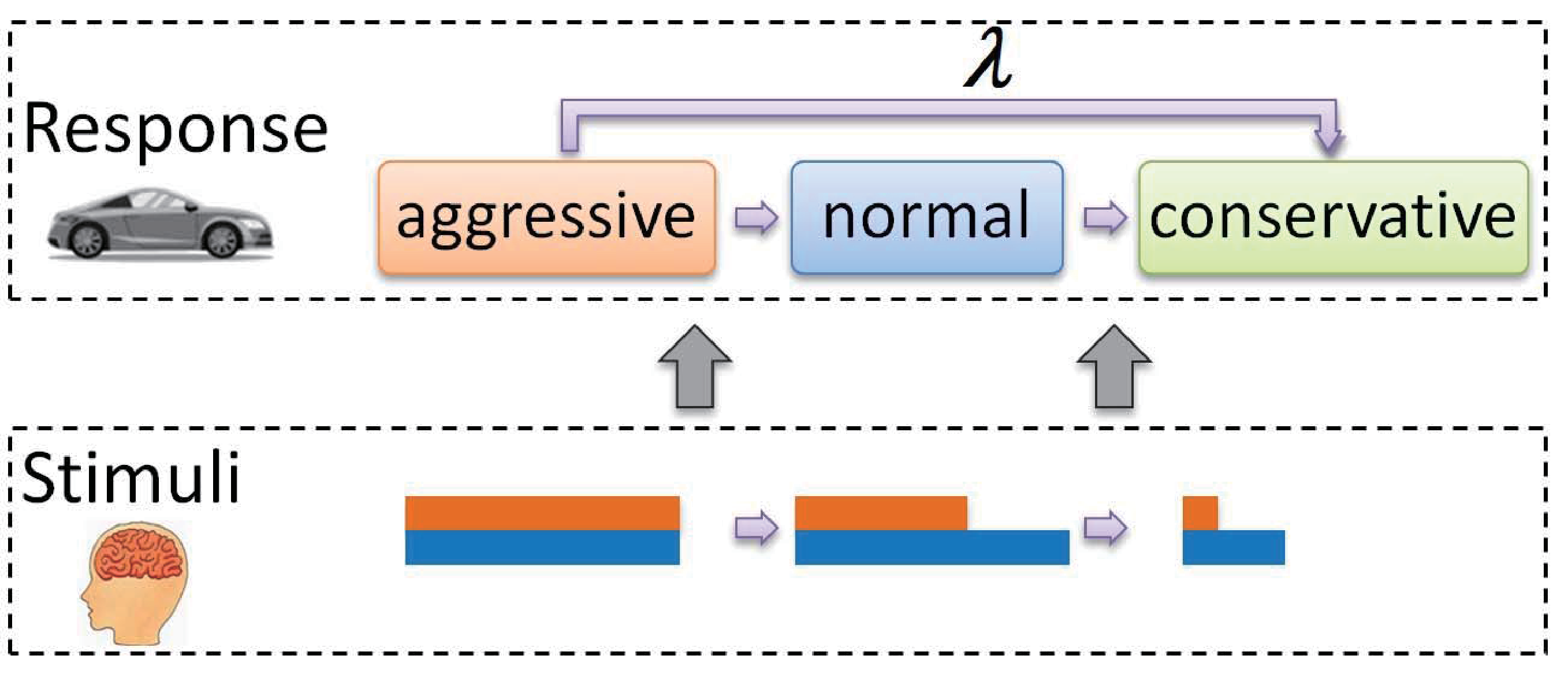}}
\caption{\label{fig-cons}
The relationship between traffic resources / driving credits and  driving behaviors in our CATS framework. }
\end{figure}

 We use the emotion-based driving behavior control model to determine the parameter vector. In the real world, different drivers have different standards on what is considered to be a safe distance, the adequate conditions for lane changing, and so on. Individual drivers may alter their driving preferences as driving credit and road resources change. For example, when driving credit is low, drivers may become more prudent while driving. In order to represent the influence of an individual's emotion on driving behavior, we class three typical driving behavior types for the dri-vehicles using the personalized modeling method based on theories of transportation psychology: conservative, normal, and aggressive driving behavior. Conservative driving refers to a driving temperament where drivers strictly follow traffic rules strictly and is hardly influenced by traffic resource and driving credit. Normal driving and aggressive driving tend to be conservative when traffic resources and driving credit are low (Figure \ref{fig-cons}). We use the Weber-Fechner law to simulate the relationship between driving preference and traffic resource / driving credit in the CATS framework.

 The Weber-Fechner law is used to obtain the movement parameters for the three driving behavior types mentioned previously. Vector $A_0$, $B_0$, $C_0$ represent values of the parameter vector $\left( {{v_0},T,{a_{\max }},{b_{\mathrm{com}}},\delta ,{s_0},\eta ,{\mu _1},{\mu _2},{\mu _3},\Delta \tau } \right)$
corresponding to the three driving behavior types at the initial time, i.e. the time when traffic resource and driving credit values are allocated. The Weber-Fechner law is used to model the relationship between the parameters and traffic resources / driving credit.

\begin{equation}\label{eq51}
A\left( {t,i} \right) = {A_0}
\end{equation}

\begin{equation}\label{eq52}
B\left( {t,i} \right) = \left( {{B_0} - {A_0}} \right)\lambda \left( {t,i} \right) + {A_0}
\end{equation}

\begin{equation}\label{eq53}
C\left( {t,i} \right) = \left( {{C_0} - {A_0}} \right)\lambda \left( {t,i} \right) + {A_0}
\end{equation}

$A\left( {t,i} \right)$, $B\left( {t,i} \right)$ and $C\left( {t,i} \right)$ represent the value of the parameter vector $\left( {{v_0},T,{a_{\max }},{b_{\mathrm{com}}},\delta ,{s_0},\eta ,{\mu _1},{\mu _2},{\mu _3},\Delta \tau } \right)$ when the driving tendency type of the $i$th dri-vehicle is conservative, normal, or aggressive, respectively. $\lambda\left( {t,i} \right)$  is the intensity value obtained using the Weber-Fechner law:
\[\lambda \left( {t,i} \right) = \frac{{\ln \left( {1 + \sigma \left( {t,i} \right)} \right)}}{{\ln 2}}\]

Where $\sigma \left( {t,i} \right) $ represents the normalized driving credit and traffic resource values of dri-vehicle $i$ at the time of $t$ .
\[\sigma \left( {t,i} \right) = \min \left( {\frac{{{\ell _i}\left( t \right)}}{{{\ell _0}}},\frac{{{\wp _i}\left( t \right) - {\wp _{\max }}}}{{{\wp _0} - {\wp _{\max }}}}} \right)\]

\section{DISCUSSION}
Now we can model the movements of vehicles in the CATS simulation. There are some problems we must discuss here.

Firstly, in relation to traffic congestion, traffic violation reporters will be rewarded with certain driving resource values under the CATS framework. Since the driving resource value is an important parameter related to the ability of dri-vehicles to endure traffic jams, the following phenomena may occur: as the value increases, some dri-vehicles are likely have to pay more resource values for congestion costs and eliminate the need to bypass bottlenecks, thereby aggravating congestion. In fact, a dri-vehicle's traffic resources and driving credit will be reduced due to traffic violations under the CATS framework. In order to maintain the right to drive, dri-vehicles will tend to drive more prudently and attempt to avoid the continuous occurrence of traffic violations, making it more difficult for other dri-vehicles to obtain resource values. Therefore, the amount of transportation resource value that can be obtained is limited. In addition, the conservative type dri-vehicles are likely to avoid congested roads during the process of path planning, regardless of how much resource value they possess. In this article, we only consider the impact of the CATS on traffic behavior types. Another study related to conducting the path planning is currently underway.

Secondly, the traffic accident rate is an important indicator of traffic safety. In general, dri-vehicles with aggressive driving behavior have a higher probability of committing traffic violations and being involved in traffic accidents, when compared to those with conservative driving tendencies. Under the CATS framework, once a dri-vehicle breaks a traffic rule, both its traffic resource value and traffic credit value will be reduced as a result. Consequently, the dri-vehicle's behavior will become more conservative, according to Equation (\ref{eq52}) and Equation (\ref{eq53}). Therefore, the CATS will reduce the rate of traffic accident occurrence.

Finally, the CATS framework only provides a strategy for lane-changing decision-making. The detailed process of lane-changing can be worked out using the method provided by Wang et al. \cite{Wang2014a}.

\section{RESULTS}

\subsection{Traffic Evolutions in the CATS Simulation}

\subsubsection{Relationship between Driving Behavior and Number of Violations}

\begin{figure}[h]
\centerline{\includegraphics[width=3.0in]{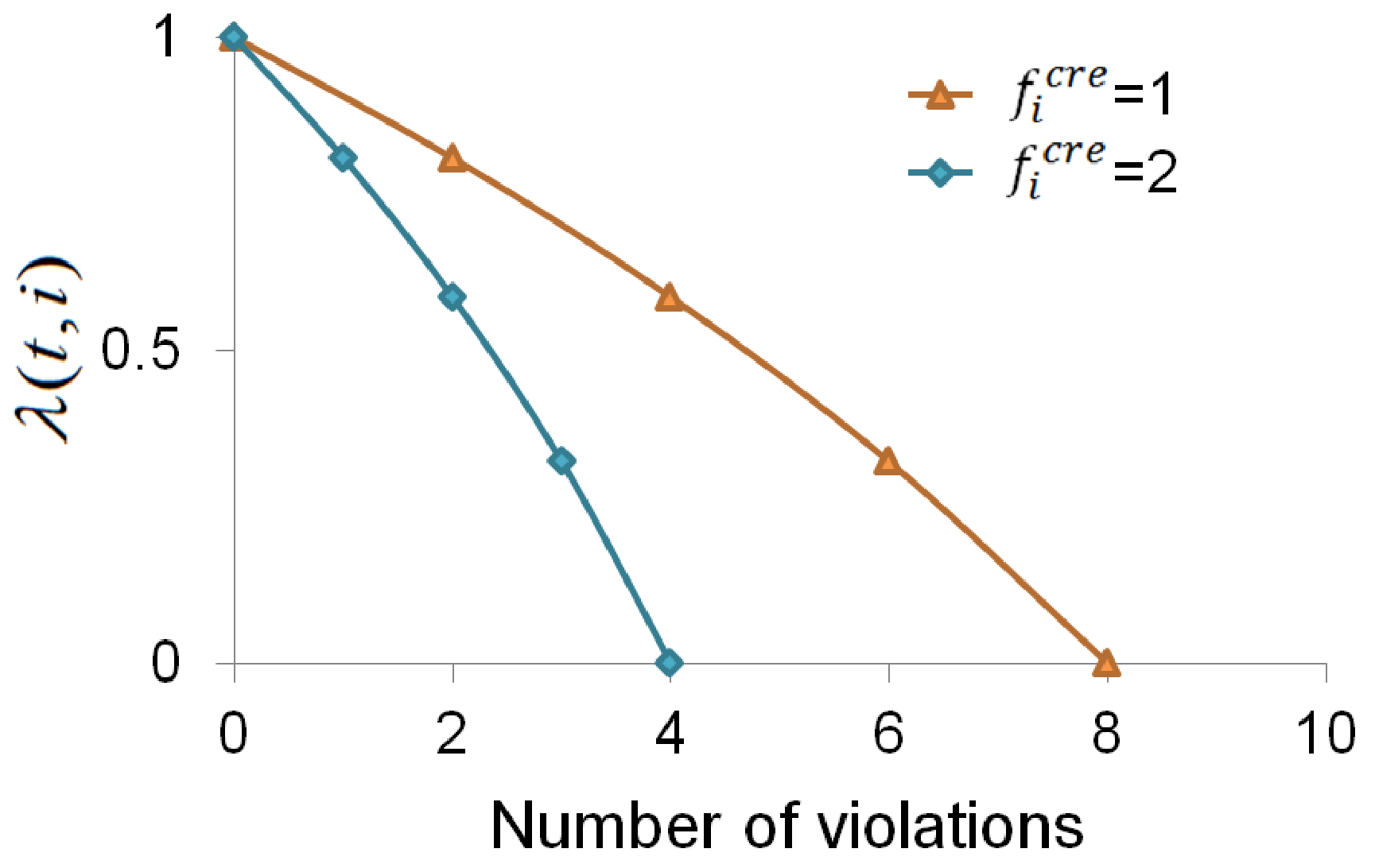}}
\caption{\label{fig-one} Relationships between driving behaviors and computing time. }
\end{figure}

In this part, we discuss the relationship between driving behaviors and computing time. According to Equation (\ref{eq51}), Equation (\ref{eq52}) and Equation (\ref{eq53}), for a dri-vehicle $i$, its driving behavior is not suddenly changed. It is determined by $\lambda \left( {t,i} \right)$ . It changes from its initial driving behavior to conservative driving when $\lambda \left( {t,i} \right)$  changes from 1 to 0. Figure \ref{fig-one} shows how $\lambda \left( {t,i} \right)$ changes as the number of violations increases for dri-vehicle $i$.
In this experiment, we let ${\wp _0} = 10 , {\wp ^{\min }} = 2 , {\ell _0} = 10$ . If a dri-vehicle drives against a traffic regulation, we let $f_i^{\mathrm{cre}} = f_j^{\mathrm{cre}} = f_i^{\mathrm{res}} = f_j^{\mathrm{res}} = 2 $  ($ \forall i,j \in \Lambda $) in order to simplify experiments. It shows that dri-vehicles drive more conservative when the number of violations increases. The greater the cost of one violation, the more conservative the dri-vehicles are likely to be.

\subsubsection{Relationship between Desired Velocity and Number of Violations}

\begin{figure}[h]
\centerline{\includegraphics[width=3.0in]{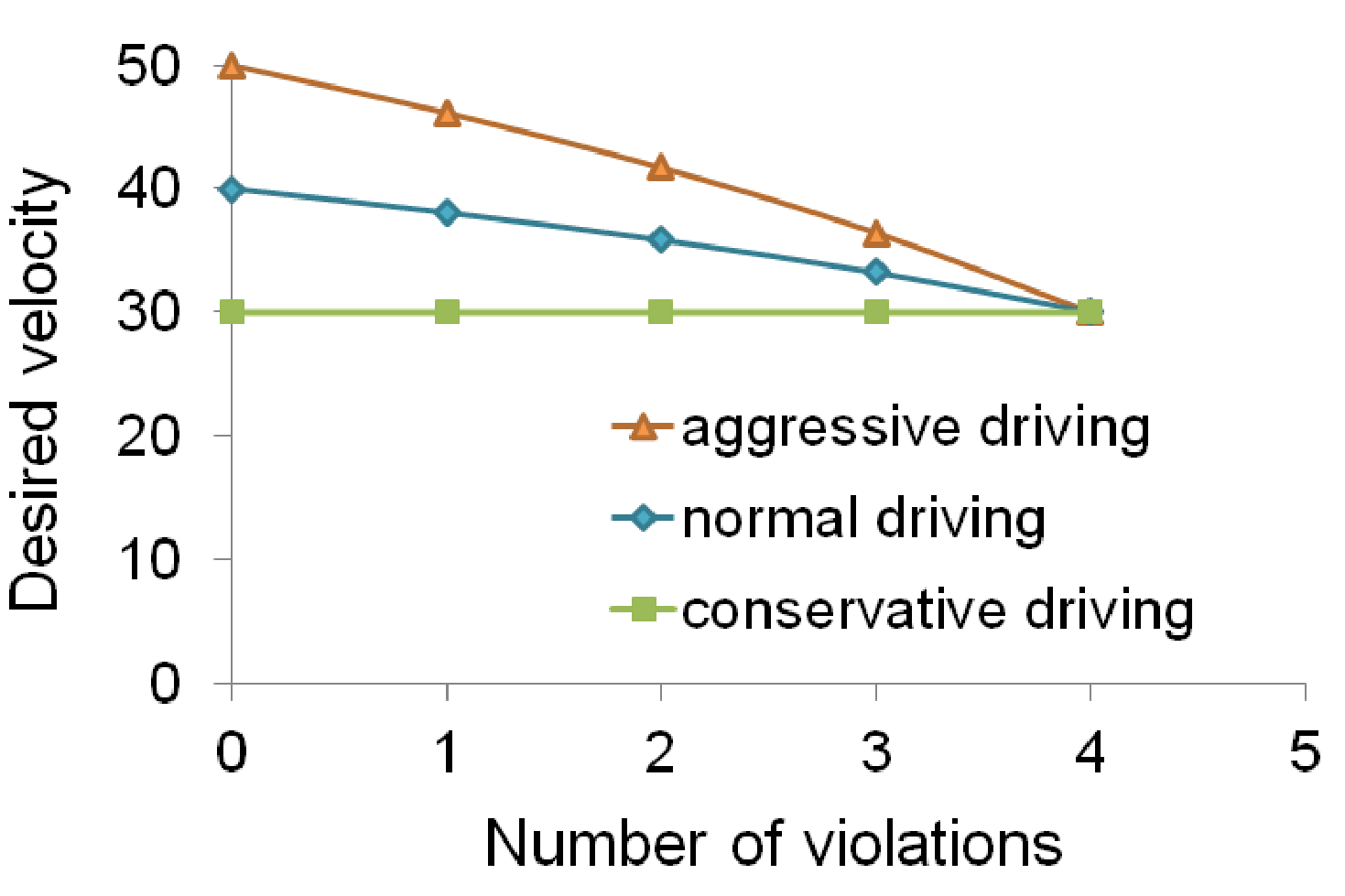}}
\caption{\label{fig-two} Relationships between desired velocity and number of violations.}
\end{figure}

Here we discuss how $v_0$ changes when the number of violations increases for a dri-vehicle. The situation is described as: A downtown traffic simulation and three driving behavior types (conservative, normal and aggressive) exist. We let initial values of  $v_0$ for the three types of dri-vehicles are 50km/h, 40km/h and 30km/h, respectively. ${\wp _0} = 10,{\wp^{\min }} = 2,{\ell _0} = 10$,  If a dri-vehicle drives againsts a traffic regulation, $f_i^{\mathrm{cre}} = f_j^{\mathrm{cre}} = f_i^{\mathrm{res}} = f_j^{\mathrm{res}} = 2$  $\forall i,j \in \Lambda $ in order to simplify experiments. Figure \ref{fig-two} shows how $v_0$ changes when the number of violations increases for the three types of dri-vehicles.

\subsubsection{Relationship between Lane-changing Duration and Number of Violations}

\begin{figure}[h]
\centerline{\includegraphics[width=3.0in]{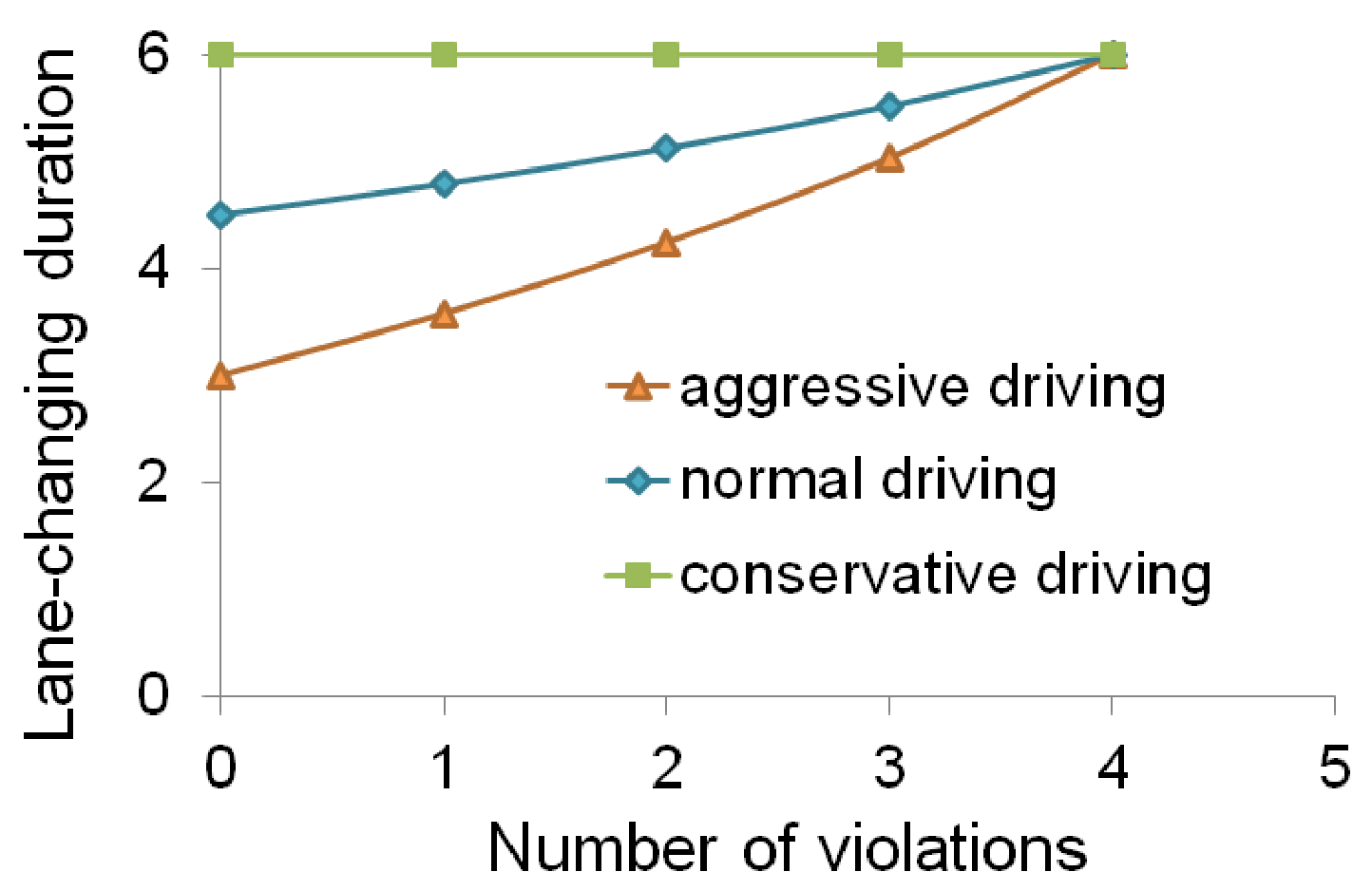}}
\caption{\label{fig-three} Relationships between Lane-changing duration and number of violations. }
\end{figure}

Here we discuss how $\Delta\tau$ changes when the number of violations increases for a dri-vehicle. The situation is described as: A downtown traffic simulation and three driving behavior types (conservative, normal and aggressive) exist. We let initial values of $\Delta\tau$ for the three types of dri-vehicles are 3s, 5s and 6s, respectively. ${\wp _0} = 10,{\wp ^{\min }} = 2,{\ell _0} = 10$ ,  If a dri-vehicle drives againsts a traffic regulation, $f_i^{\mathrm{cre}} = f_j^{\mathrm{cre}} = f_i^{\mathrm{res}} = f_j^{\mathrm{res}} = 2$  $\forall i,j \in \Lambda $ in order to simplify experiments. Figure \ref{fig-three} shows how $\Delta\tau$ changes when the number of violations increases for these three types of dri-vehicles.

\subsection{Simulation and Comparison}
In this section, we will discuss how traffic accident rates change in the CATS framework.

The simulation we conduct involves 2,000 vehicles, among which 25\% are classed as conservative driving type, while 50\% are classed as normal driving type, and 25\% as aggressive driving type.

Considering that traffic cameras are highly visible since they are installed outdoors and that most commercial navigation applications alert users when vehicles are in close proximity to a traffic camera, the following assumptions in relation to the traffic violation rates at the areas covered by traffic cameras are made:
dri-vehicles with an aggressive driving behavior type: 2\% (i.e. 2 out of every 100 dri-vehicles will break traffic rules per day);
dri-vehicles with a normal driving behavior type: 1\%;
and dri-vehicles with a conservative driving behavior type: 0\%.
For areas not covered by traffic cameras, the following assumptions in relation to the traffic violation rates are made:
dri-vehicles with an aggressive driving behavior type: 10\% (i.e. 10 out of every 100 dri-vehicles will break traffic rules per day);
dri-vehicles with a normal driving behavior type: 5\%;
and dri-vehicles with a conservative driving behavior type: 0\%.

Let the violation recurrence rate be 50\%, which means that 50\% of the violations that occur each day are not the first instance of occurrence. To simplify the simulation, if a dri-vehicle chooses to travel to a location within a congested area, he will be deducted two units of traffic resources. Additionally, for any traffic violation, the dri-vehicle will be deducted two units of drive credit and two units of traffic resources. It is also assumed that two nearby cars will detect the same violation and report it at the same time. Furthermore, it is assumed that the value of driving parameters corresponding to conservative, normal, and aggressive driving is ${B_0} = 0.5 * ({A_0} + {C_0})$.

\begin{figure}[h]
\centerline{\includegraphics[width=3.0in]{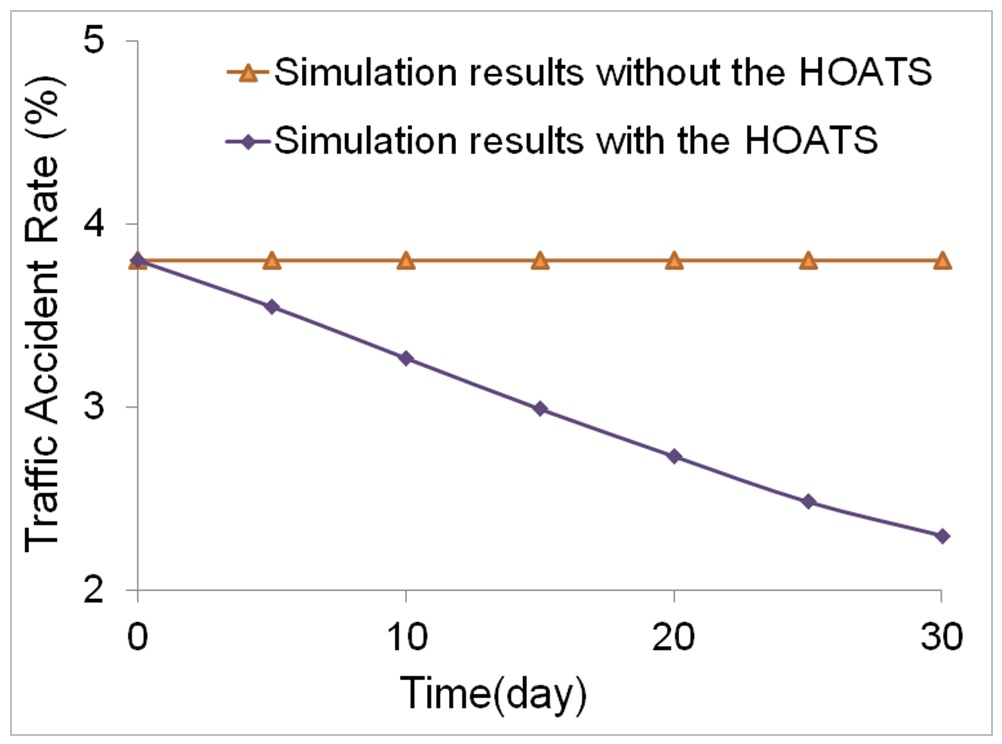}}
\caption{\label{fig-withandwithoutHOATS} How traffic accident rates changed over time in the scenarios that
are based and not based on the CATSS.}
\end{figure}

Figure \ref{fig-withandwithoutHOATS} shows how traffic accident rates change over time in scenarios that are based on the CATS and those which are not. It is clear that the CATS can significantly reduce traffic accidents.

\begin{figure}[h]
\centerline{\includegraphics[width=3.0in]{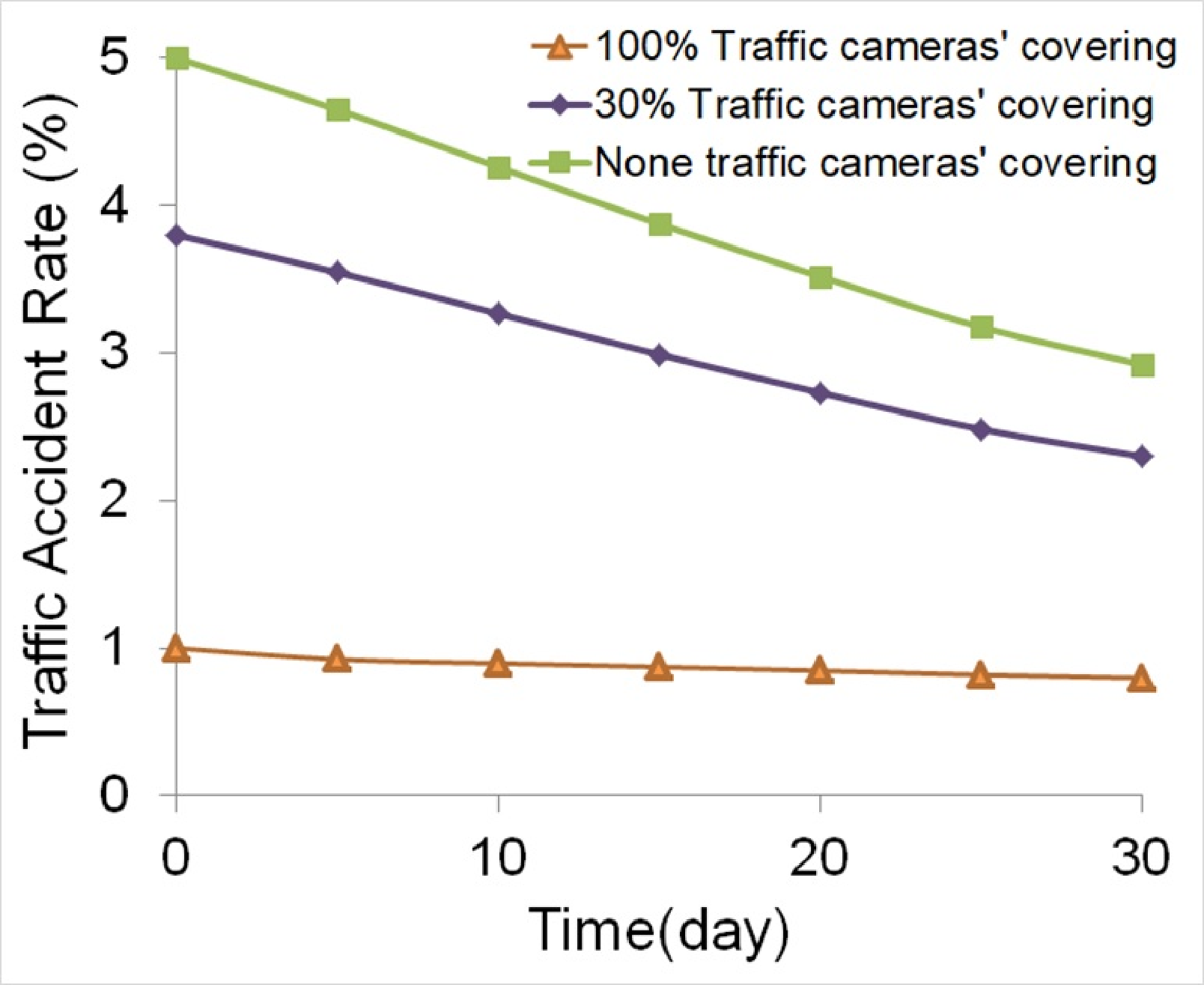}}
\caption{\label{fig-traffic accident rate-time} How traffic accident rates changed over time in different scenarios in the CATS simulation.}
\end{figure}

Figure \ref{fig-traffic accident rate-time} presents how traffic accident rates change over time in three scenarios ( no camera coverage, 30\% camera coverage, and 100\% camera coverage) in the CATS simulation.

It is demonstrated that the accident rate decreases over time in all scenarios. In the scenario with 100\% camera coverage, the traffic accident rate only drops minimally. This is because the initial accident rate is very low in this scenario when compared to other scenarios.

\begin{figure*}[t]
\centerline{\includegraphics[width=7.0in]{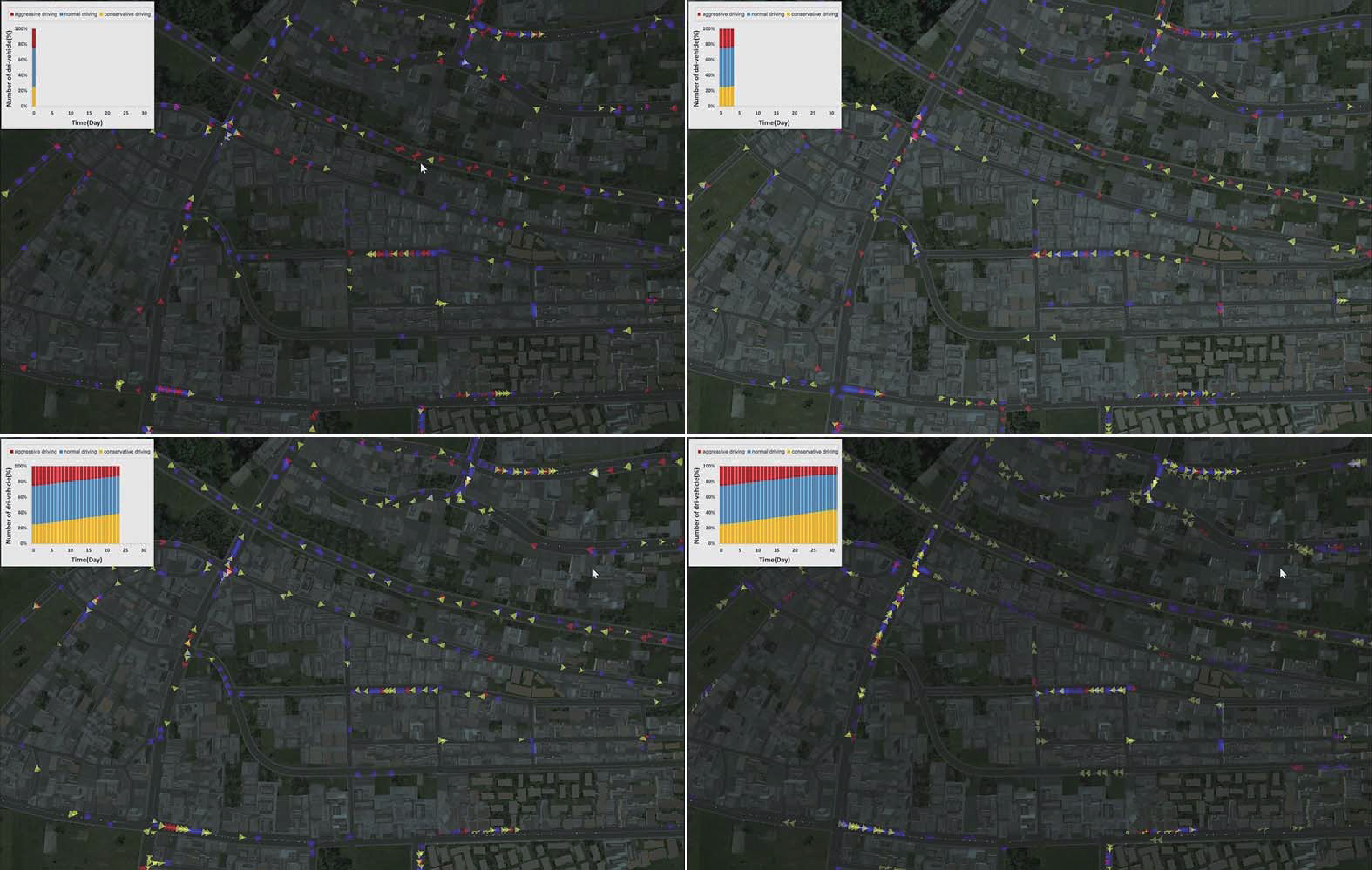}}
\caption{\label{fig-sim}Snapshots of traffic animation results about how the percentage of dri-vehicles of the three driving behavior types changed over time.}
\end{figure*}

\begin{figure}[h]
\centerline{\includegraphics[width=3.0in]{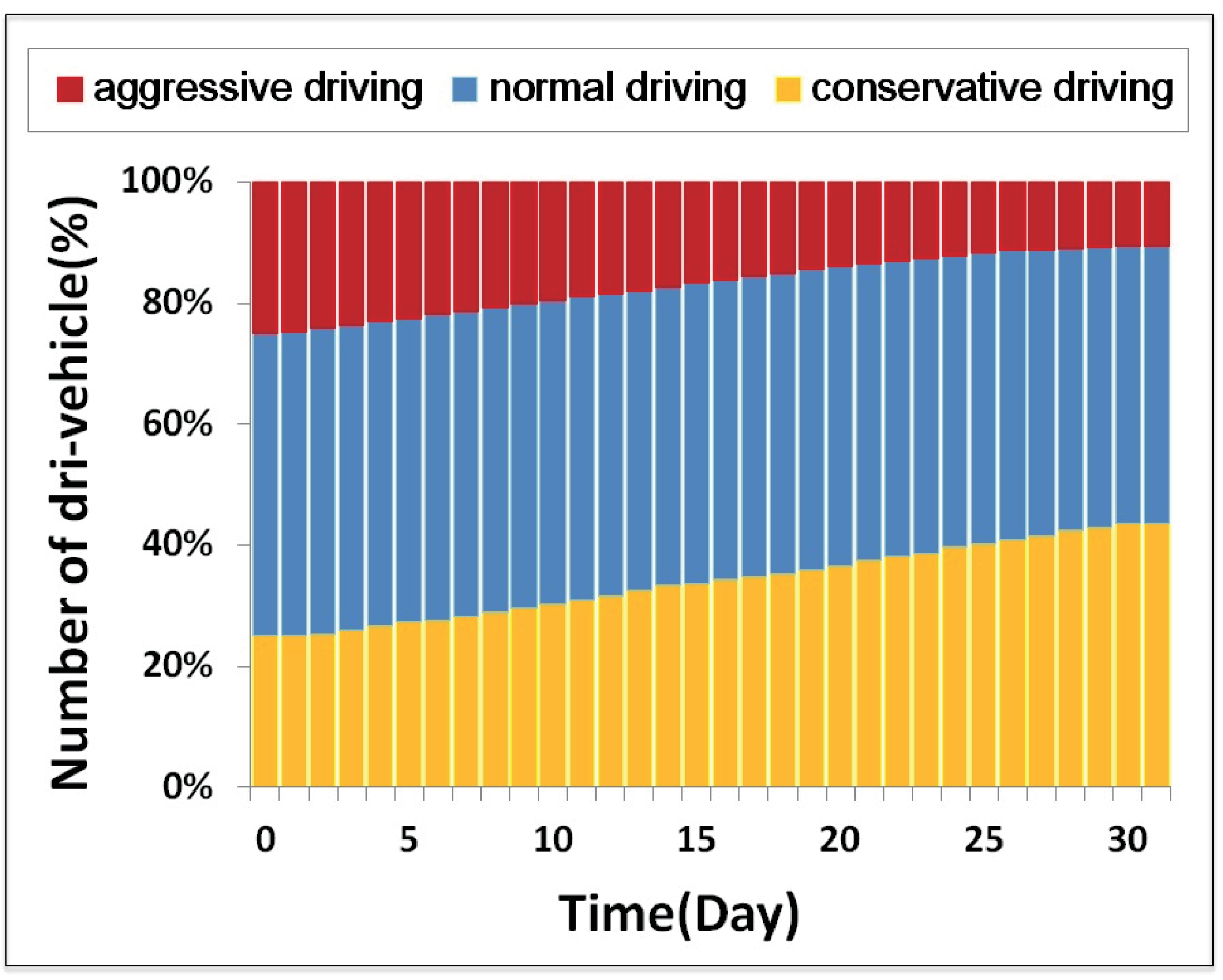}}
\caption{\label{fig-number of car-time}How the percentage of dri-vehicles of the three driving behavior types changed over time when the scenario with 30\% camera coverage was simulated.}
\end{figure}

Figure \ref{fig-number of car-time} shows how the percentage of the three described driving behavior types changes over time when the 30\% camera coverage scenario is simulated using the CATS. As shown, with the passage of time, the number of dri-vehicles with a conservative driving behavior type increases steadily while the number of dri-vehicles with an aggressive driving behavior type decreases gradually. The number of dri-vehicles with a normal driving behavior type increases minimally during the early stage of simulation and then decreases gradually later. The reason for the increase is that aggressive dri-vehicles become normal dri-vehicles, and then become conservative dri-vehicles. Figure \ref{fig-sim} shows snapshots of the traffic animation results.

\subsection{Comparison of Performance Scaling}

To better evaluate the performance of CATS simulation, we compare the computational time of it to the SUMO, which is an agent-based traffic simulation system designed to handle large road networks \cite{SUMO}. The results are collected on an Intel Core(TM) i7-6700HQ 2.60GHZ CPU with an independent graphics card.

\begin{figure}[h]
\centerline{\includegraphics[width=3.0in]{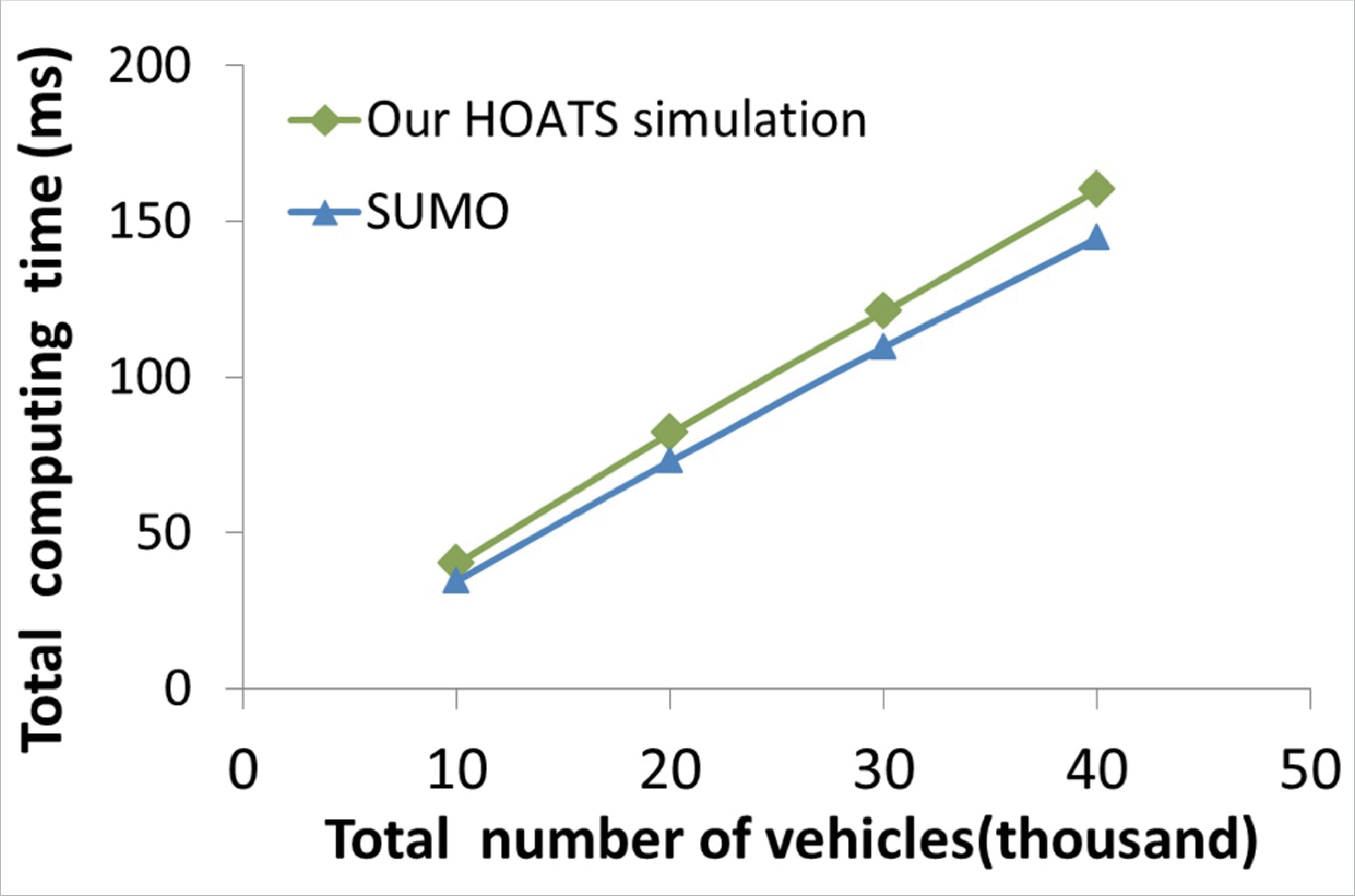}}
\caption{\label{fig-performance} Comparison of the per scaling of the SUMO and our CATS simulation as the number of vehicles increases.}
\end{figure}

Results are shown in Figure \ref{fig-performance}. It is clear that the CATS simulation has nearly the same efficiency as that of the SUMO.

\section{CONCLUSIONS}

In this paper, we have presented an innovative traffic management framework named CATS framework. Through creatively introducing mutual supervision and a reward/punishment mechanism, the CATS framework is able to solve problems in existing urban traffic systems. In addition, we have also provided a traffic simulation method to model and deduce the framework. We have introduced how to transform the above mechanisms into traffic simulations, in which traffic economics, psychology and dynamic traffic simulations are combined. Simulation results demonstrate that the CATS framework can significantly reduce the number of accidents. In the future, we hope to explore modeling of path planning within the CATS framework.

\section*{ACKNOWLEDGEMENTS}
This work is supported and funded by the National Natural Science Foundation of China
(Grant No. 61602425, 61472370, 61672469, 61602420, 61502433), the Open Project Program of the State Key Lab of CAD\&CG (Grant No. A1717), Zhejiang University. We would like to thank the reviewers for their constructive comments and suggestions.



\bibliographystyle{unsrt}
\bibliography{traffic}

\end{document}